\documentclass{article}
\usepackage{spconf,amsmath,graphicx}
\usepackage{amssymb}
\usepackage{bm}
\usepackage{booktabs}
\usepackage{standalone}
\usepackage{pgfplots}
\usepackage{balance}
\usepackage{tikz}
\usetikzlibrary{calc}
\makeatletter
\newcommand{\gettikzxy}[3]{%
  \tikz@scan@one@point\pgfutil@firstofone#1\relax
  \edef#2{\the\pgf@x}%
  \edef#3{\the\pgf@y}%
}

\usepackage{caption}
\usetikzlibrary{spy,backgrounds}
\pgfplotsset{compat=newest}
\usetikzlibrary{plotmarks}
\usetikzlibrary{arrows.meta}

\title{Active Inference Framework for Closed-Loop Sensing, Communication, and Control in UAV Systems}

\name{
Guangjin Pan\textsuperscript{*},  Liping Bai\textsuperscript{*}, Zhuojun Tian\textsuperscript{†}, Hui Chen\textsuperscript{*}, Mehdi Bennis\textsuperscript{†}, Henk Wymeersch\textsuperscript{*}}
\address{\textsuperscript{*}Department of Electrical Engineering, Chalmers University of Technology, Sweden \\
\textsuperscript{†} Centre of Wireless Communications,  University of Oulu, Finland \\}
\begin{document}
%
    \maketitle
\begin{abstract}
Integrated sensing and communication (ISAC) is a core technology for 6G, and its application to closed-loop sensing, communication, and control (SCC) enables various services. Existing SCC solutions often treat sensing and control separately, leading to suboptimal performance and resource usage. In this work, we introduce the active inference framework (AIF) into SCC-enabled unmanned aerial vehicle (UAV) systems for joint state estimation, control, and sensing resource allocation. By formulating a unified generative model, the problem reduces to minimizing variational free energy for inference and expected free energy for action planning. Simulation results show that both control cost and sensing cost are reduced relative to baselines.
\end{abstract}
\vspace{-1mm}
\section{Introduction}
\vspace{-2mm}

Integrated sensing and communication (ISAC) is envisioned as a key enabler of 6G networks. 
Building on this, closed-loop ISAC systems that jointly consider sensing, communication, and control (SCC) show strong potential in applications such as autonomous driving~\cite{Wang_AutonomousDriving_2019}, collaborative robotics~\cite{Goal-Oriented_robot}, and unmanned aerial vehicles (UAVs)~\cite{Chang_IntegratedUAV_2022}. By exploiting wireless system for both sensing and coordination, SCC can enable highly adaptive and intelligent networked control.

However, most existing SCC solutions address sensing, communication, and control in a decoupled manner, which often leads to performance degradation and inefficient resource utilization. Focusing on sensing and communication, ISAC resource allocation research has primarily aimed at improving sensing performance~\cite{li2024maximizing, li2025massive} or balancing sensing and communication~\cite{khalili2024efficient, zou2024energy}, while neglecting control objectives. Some studies incorporate sensing feedback into control, but they typically assume fixed sensing quality and therefore overlook the fundamental trade-off between estimation accuracy and sensing overhead~\cite{duan2022distributed, jin2025co}. A few recent works have started to consider SCC systems \cite{meng2024communication, fang2024sensing1,fang2025sensing2,pan2025rate}. For instance, \cite{meng2024communication} studies communication–control coupling in UAV networks, while \cite{fang2024sensing1, fang2025sensing2} focus on communication and computing resource allocation. In \cite{pan2025rate}, the impact of observation compression on closed-loop ISAC systems is investigated. Additionaaly, the authors in \cite{10446575} use AIF for UAV planning, but does not incorporate control or sensing. However, there is still a lack of a comprehensive approach that integrates state estimation, control, and sensing resource management, which is essential for realizing intelligent closed-loop ISAC systems.

The active inference framework (AIF), a Bayesian brain-inspired paradigm, simultaneously abstracts the generative model of the environment and infers Bayes-optimal behavior by minimizing free energy functionals. As a formal framework to jointly optimize perception (sensing) and action (communication and control),  we apply AIF into SCC where estimation, control, and sensing resource allocation are inherently coupled~\cite{obite2023active, krayani2023goal}. In this work, we present the first application of the AIF \cite{krayani2023goal, obite2023active, otoshi2025coordinated}
to closed-loop ISAC systems, where the SCC problem is formulated into a free-energy minimization problem under the AIF. Specifically, by constructing a unified generative model that incorporates UAV dynamics, sensing observations, and wireless resource allocation, variational free energy (VFE) is minimized to infer past states, while expected free energy (EFE) is minimized to plan future control actions and allocate resources. The proposed AIF provides a principled approach to the joint design of control and sensing resource management, opening a new paradigm for intelligent SCC system design.

\color{black}

\begin{figure}[tb]
    \centering
    \includegraphics[width=0.75\linewidth]{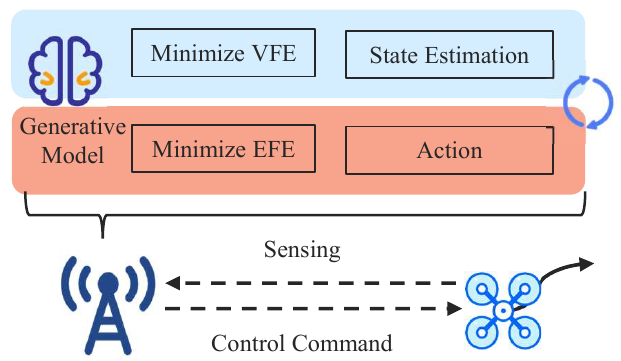}
    \caption{UAV control system with a sensing BS under the AIF.}
    \label{fig:system}
\end{figure}

\vspace{-2mm}
\section{System model}
\vspace{-2mm}

As illustrated in Fig.~\ref{fig:system}, we consider a UAV control system assisted by a sensing base station (BS). The BS exploits the wireless channels to localize the UAV, and based on the sensing results, performs both UAV control and sensing resource allocation for the next step. In this section, we first present the UAV state transition model and the wireless sensing model, followed by the objective model that captures the trade-off between the control cost and sensing cost. Finally, these models are unified under the AIF framework.

\vspace{-3mm}
\subsection{State Transition Model}
\vspace{-2mm}

\label{subsec:state}
The UAV control system is assisted by a sensing BS located at the origin of a polar coordinate system, under a MIMO-OFDM setup. The UAV state at time $t$ is $x_t \triangleq [\rho_t , \psi_t , v_{\rho,t} , \omega_t]^\top$ where $\rho_t$ is the radial distance to the BS, $\psi_t$ is the azimuth angle, $v_{\rho,t}$ is the radial velocity, and $\omega_t$ is the angular velocity. With a sampling interval $\Delta t$, the UAV dynamics follow a constant-acceleration model: $x_{t+1} = A x_t + B u_t + n_t^x$,
with system matrices $    A =
    \begin{bmatrix}
        1 & 0 & \Delta t & 0 \\
        0 & 1 & 0 & \Delta t \\
        0 & 0 & 1 & 0 \\
        0 & 0 & 0 & 1
    \end{bmatrix}$ and $  B =   \begin{bmatrix}
        \tfrac{1}{2}\Delta t^2 & 0 \\
        0 & \tfrac{1}{2}\Delta t^2 \\
        \Delta t & 0 \\
        0 & \Delta t
    \end{bmatrix}.$
The control input $u_t = [a_{\rho,t}, a_{\omega,t}]^\top$ consists of radial and angular accelerations. The process noise $n_t^x \sim \mathcal{N}(0, {Q})$ has covariance ${Q}$.

\vspace{-3mm}
\subsection{Sensing Model}
\vspace{-2mm}

We consider wireless localization based on the UAV uplink channel, which serves as a representative case of wireless sensing \cite{gonzalez2024integrated, pan2025ai}. The BS is equipped with $K$ subcarriers over a total bandwidth $B$ with subcarrier spacing $\Delta f = B/K$. At each time slot $t$, the BS allocates $k_t \leq K$ consecutive subcarriers in the frequency domain for UAV localization.
To simplify the model, we use the Cramér--Rao lower bound (CRLB) instead of the localization algorithm. The TOA-induced ranging error variance using $k_t$ subcarriers is based on CRLB for the TOA estimation, which can be expressed as $
    \sigma_{r,t}^2(k_t) = c^2 \mathrm{var}(\hat{\tau}_t) =
    {c^2}/{(8 \pi^2 \, \gamma_t \, \frac{k_t^2}{K^2} B^2)}$~\cite{TOA-CRLB},
where $c$ is the speed of light. For AOA estimation, we assume a constant error variance $\sigma_\theta^2$, as the CRLB becomes nearly bandwidth-independent when the bandwidth is much smaller than the carrier frequency \cite{AOA-CRLB}. Therefore, The observation vector at time $t$ is approximated by a linear Gaussian model, i.e., 
$y_t \sim \mathcal{N}(y_t;C x_t, R(k_t))$, with measurement matrix $C = \begin{bmatrix} {I}_2, {0}_{2\times 2} \end{bmatrix}$ and covariance ${R}(k_t) = \mathrm{diag}\!\big( \sigma_{r,t}^2(k_t),\, \sigma_\theta^2 \big)$.

\vspace{-4mm}
\subsection{Objective Model}\label{Sec_2_3}
\vspace{-2mm}

Given the desired trajectory $x_t^{\text{desired}}$, the state deviation is 
$\tilde{x}_t = x_t - x_t^{\text{desired}}$. 
The control cost can be expressed as $J_t^{\text{control}} 
    = \tilde{x}_t^\top Q_{\text{goal}} \tilde{x}_t 
    + u_t^\top R_{\text{goal}} u_t$,
where $Q_{\text{goal}} \succ 0$ penalizes state deviation and 
$R_{\text{goal}} \succ 0$ penalizes control effort. They reflect control energy consumption or actuator limitations. For wireless localization, allocating more subcarriers for sensing consumes extra computational and power resources at the BS. We model the sensing cost as a function of the number of allocated subcarriers $k_t$, i.e., $J_t^{\text{sensing}} = \alpha  k_t^2 + \beta \log\det(R(k_t))$,
where $\alpha>0$ penalizes quadratic subcarrier usage and $\beta>0$ weights the uncertainty term $\log\det(R(k_t))$ induced by the covariance matrix approximated by the CRLB. The overall objective is to balance control performance with sensing resource consumption by minimizing the expected cumulative cost: $J = \mathbb{E} \left[ \sum_{t=0}^{T} 
    \big( J_t^{\text{control}} + J_t^{\text{sensing}} \big) \right]$,
where the trade-off between control and sensing cost can be adjusted by changing $\alpha$ in $J_t^{\text{sensing}}$. In the AIF, this objective model does not serve as a direct optimization target; instead, it defines the agent’s \emph{goal prior}, i.e., a prior distribution that encodes the agent’s preferences over future states, observations, and actions, which guides decision-making via free-energy minimization \cite{van2021application}.

\vspace{-4mm}
\subsection{Problem Formulation via AIF}
\vspace{-2mm}

Building upon the above system model, we formulate the UAV control and sensing co-design problem within the framework of active inference. The generative model of the agent is characterized by three main components:
\begin{itemize}
    \vspace{-2mm}
    \item \textbf{State:}  The UAV state evolves as a linear Gaussian process $
        P(x_t \mid x_{t-1}, u_{t-1}) 
        = \mathcal{N}\!\big(A x_{t-1} + B u_{t-1}, {Q}\big)$.
        
    \vspace{-2mm}
    \item \textbf{Observation:}  Based on $k_t$ subcarriers from the UAV uplink, the BS obtains $P(y_t \! \mid\! x_t, k_t) = \mathcal{N}\!\big(Cx_t, {R}(k_t)\big) $,
    where ${R}(k_t)$ captures the TOA and AOA variances.
    \vspace{-2mm}
    \item \textbf{Action:}  
    Action is defined as $a_t = (u_t, k_{t+1})$ where control action $u_t$ represents the UAV control input, and the sensing action $k_{t+1}$ specifies the number of subcarriers allocated for localization at the next time step.
    \vspace{-2mm}
\end{itemize}

In this work, we consider the single-step inference case. For each time slot $t$, the generative model conditioned on the action sequence $a_t$ can be written as~\cite{van2021application}:
\vspace{-3mm}
\begin{align}
    P_g(x_{t}, y_{t} \! \! \mid \! \! a_{t}) 
    &=  P_{g,1}(x_{t}, {y}_{t} \! \mid \!\!  {u}_{t-1}, k_{t})   P_{g,2}(x_{t}, {y}_{t} \! \mid \! x_t, {u}_{t-1}, k_{t})\, \nonumber \\
    &   \times  \tilde{P}(x_{t}, {y}_{t}, {u}_{t-1}, k_{t+1})  \\
    & = P(x_1) 
       \! \prod_{t=2}^t \! P(x_t \!\mid\! x_{t-1},\! {u}_{t-1}) 
       \!\prod_{t=1}^t\! P({y}_t \!\mid\! x_t,\! k_t) \nonumber \\
    & \times
       \!\prod_{t=t+1}^T\! P(x_t \!\mid\! x_{t-1}\!,\! {u}_{t-1})
       \!\prod_{t=t+1}^T\!  P({y}_t \!\mid\! x_t\!,\! k_t) \nonumber\\
     & \times \!\prod_{t=t+1}^T \!\tilde{P}(x_t, y_t, u_{t-1}, k_t).\label{eq_Sec2_2}
     \vspace{-1mm}
\end{align}
In \eqref{eq_Sec2_2}, the first term specifies the prior distribution over the initial state.  
The second and third terms correspond to the state transition and observation likelihoods up to the current time $t$, which collectively form 
$P_{g,1}(x_{1:t}, y_{1:t} \mid u_{1:t-1}, k_{1:t})$.  
The fourth and fifth terms constitute $P_{g,2}(x_{t+1:T}, y_{t+1:T} \mid  x_t, u_{t:T-1}, k_{t+1:T})$, which characterizes the predicted future state dynamics and observations conditioned on the present state and planned actions.  
The last factor $\tilde{P}(x_t, y_t, u_t, k_t)$ represents the agent’s goal prior. In active inference, the goal prior encodes preferences over future states, observations, and actions to guide decision-making. In this problem, it is modeled as a trade-off between control performance and sensing cost, as discussed in Section \ref{Sec_2_3}:
\vspace{-3mm}
\begin{align}
    &\tilde{P}(x_{t+1}, y_{t+1}, u_t, k_{t+1}) \nonumber
     \\ 
     & \propto \exp \left(
        -  J_t^{\text{control}}(x_{t+1}, u_t)
        -  J_t^{\text{sensing}}(k_{t+1})
    \right).
\end{align}

Under this factorization, the free energy can be decomposed as~\cite{van2021application}
\vspace{-3mm}
\begin{align}
\mathcal{F}[q_1,q_2] 
    &= 
    \underbrace{\mathbb{E}_{q_1}\!\left[
        \log \frac{q_1(x_{t}\mid y_{t}, u_{t-1})}
                  {P_{g,1}(x_{t}, y_{t}\mid u_{t-1}, k_{t})}
    \right]}_{\smash{\triangleq \; \mathcal{F}_V[q_1]}} \notag \\[1ex]
    & +
    \underbrace{\begin{aligned}[t]
        \mathbb{E}_{q_2}\!\bigg[
            \log \frac{
                q_2(x_{t+1}, y_{t+1}\mid x_t, u_{t}, k_{t+1})
            }{
                P_{g,2}(x_{t+1}, y_{t+1}\mid x_t, u_{t}, k_{t+1})
            } \\
            - \log \tilde{P}(x_{t+1}, y_{t+1}, u_{t}, k_{t+1})
        \bigg]
    \end{aligned}}_{\smash{\triangleq \; \mathcal{G}[q_2]}} \label{eq:Free-energy}.
\end{align}
$q_1$ is the approximate posterior of past states $x_{t-1}$, while $q_2$ is the predictive distribution of future trajectories $(x_{t+1}, y_{t+1})$ under candidate actions.
Within this decomposition, the active inference agent faces two complementary tasks: 

\textbf{Inference Stage (State Estimation):} The first term in~\eqref{eq:Free-energy}, $\mathcal{F}_V[q_1]$, corresponds to the VFE. Minimizing it gives the approximate posterior: $q_1^*(x_{1:t}) = \arg\min_{q_1} \mathcal{F}_V[q_1]$,
which represents the agent’s belief over latent states given past observations and actions.

\textbf{Planning Stage (Action Optimization):}
The second term in ~\eqref{eq:Free-energy}, $\mathcal{G}[q_2]$, corresponds to the EFE. At each step $t$, the agent minimizes the EFE with respect to the predictive distribution, i.e., $q_2^* = \arg\min_{q_2} \; \mathcal{G}[q_2]$.
$q_2^*$ is the optimal predictive distribution over future states and observations under candidate actions. 
A concrete action is then extracted from which the optimal action is obtained as $(u_t^*, k_{t+1}^*) = g(q_2^*)$, where $g(\cdot)$ denotes the mapping from the predictive distribution $q_2^*$ to the corresponding action (e.g., via maximum a posteriori estimation or sampling).

\vspace{-2mm}
\section{Free Energy Minimization by Message Passing}
\vspace{-2mm}

In principle, the optimization involves the entire future horizon $(t+1):T$.  
In this work, we assume that the AIF agent has prior knowledge of the state transition and observation models, and adopt a one-step control method to simplify the problem, where only the immediate prediction of the next state and observation is considered when evaluating $\mathcal{G}[q_2]$.  

\vspace{-3mm}
\subsection{State Estimation Stage}
\vspace{-2mm}

\begin{figure}[t]
    \vspace{-2mm}
    \centering
    \begin{tikzpicture}
    \node (image) [anchor=south west]{\includegraphics[width=1\linewidth]{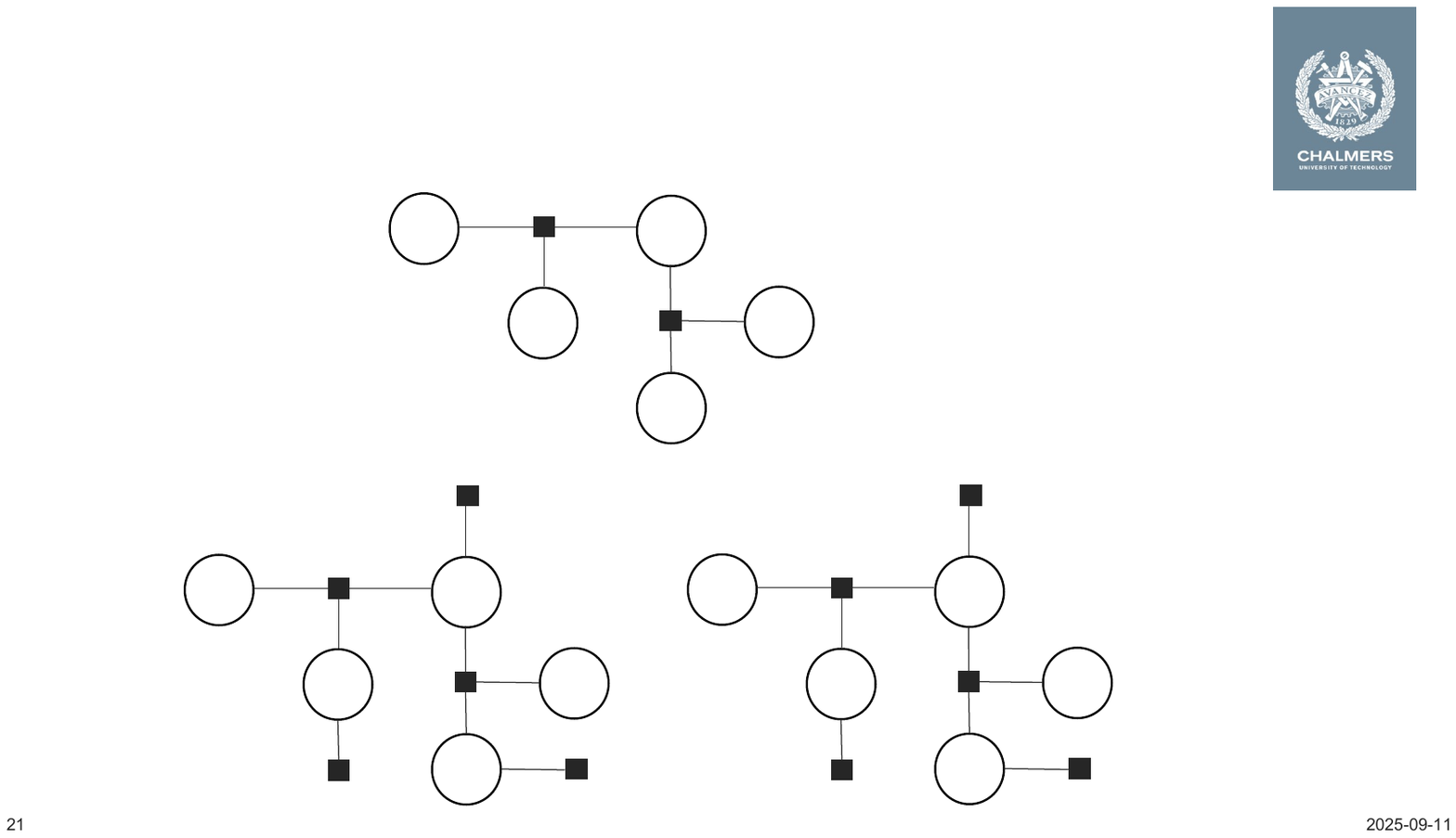}};
    \gettikzxy{(image.north east)}{\ix}{\iy};
    \node at (0.274*\ix,0.940*\iy)[rotate=0,anchor=north]{{\scriptsize $x_{t-1}$}};
    \node at (0.525*\ix,0.940*\iy)[rotate=0,anchor=north]{{\scriptsize $x_t$}};
    \node at (0.395*\ix,0.804*\iy)[rotate=0,anchor=north]{{\scriptsize $u_{t-1}$}};
    \node at (0.635*\ix,0.804*\iy)[rotate=0,anchor=north]{{\scriptsize $y_{t}$}};   
    \node at (0.525*\ix,0.686*\iy)[rotate=0,anchor=north]{{\scriptsize $k_{t}$}};
    \node at (0.395*\ix,0.995*\iy)[rotate=0,anchor=north]{{\scriptsize $f_{t}^x$}};
    \node at (0.56*\ix,0.85*\iy)[rotate=0,anchor=north]{{\scriptsize $f_{t}^y$}};

    \node at (0.495*\ix,0.61*\iy)[rotate=0,anchor=north]{{\scriptsize (a) Factor graph for state estimation}};
    \draw[gray!80, thin, ->] (0.315*\ix, 0.932*\iy) -- (0.380*\ix, 0.932*\iy);
    \draw[gray!80, thin, ->] (0.415*\ix, 0.932*\iy) -- (0.485*\ix, 0.932*\iy);
    \draw[gray!80, thin, ->] (0.38*\ix, 0.83*\iy) -- (0.38*\ix, 0.90*\iy);
    \draw[gray!80, thin, ->] (0.515*\ix, 0.805*\iy) -- (0.515*\ix, 0.85*\iy);
    \draw[gray!80, thin, ->] (0.515*\ix, 0.71*\iy) -- (0.515*\ix, 0.755*\iy);
    \draw[gray!80, thin, ->] (0.595*\ix, 0.76*\iy) -- (0.545*\ix, 0.76*\iy);

    \node at (0.0666*\ix,0.407*\iy)[rotate=0,anchor=north]{{\scriptsize $x_t$}};
    \node at (0.316*\ix,0.407*\iy)[rotate=0,anchor=north]{{\scriptsize $x_{t+1}$}};
    \node at (0.185*\ix,0.272*\iy)[rotate=0,anchor=north]{{\scriptsize $u_t$}};
    \node at (0.430*\ix,0.272*\iy)[rotate=0,anchor=north]{{\scriptsize $y_{t+1}$}};   
    \node at (0.315*\ix,0.160*\iy)[rotate=0,anchor=north]{{\scriptsize $k_{t+1}$}};

    \node at (0.19*\ix,0.49*\iy)[rotate=0,anchor=north]{{\scriptsize $f_{t+1}^x$}};
    \node at (0.35*\ix,0.335*\iy)[rotate=0,anchor=north]{{\scriptsize $f_{t+1}^y$}};
    \node at (0.13*\ix,0.17*\iy)[rotate=0,anchor=north]{{\scriptsize $f_{t}^{\text{goal},u}$}};
    \node at (0.435*\ix,0.12*\iy)[rotate=0,anchor=north]{{\scriptsize $f_{t+1}^{\text{goal},k}$}};
    \node at (0.372*\ix,0.56*\iy)[rotate=0,anchor=north]{{\scriptsize $f_{t+1}^{\text{goal},x}$}};

    \draw[gray!80, thin, ->] (0.10*\ix, 0.415*\iy) -- (0.17*\ix, 0.415*\iy);
    \draw[gray!80, thin, ->] (0.28*\ix, 0.415*\iy) -- (0.21*\ix, 0.415*\iy);
    \draw[gray!80, thin, ->] (0.205*\ix, 0.38*\iy) -- (0.205*\ix, 0.30*\iy);
    \draw[gray!80, thin, ->] (0.205*\ix, 0.14*\iy) -- (0.205*\ix, 0.19*\iy);
    \draw[gray!80, thin, ->] (0.30*\ix, 0.51*\iy) -- (0.30*\ix, 0.44*\iy);
    \draw[gray!80, thin, ->] (0.30*\ix, 0.27*\iy) -- (0.30*\ix, 0.33*\iy);

    \node at (0.255*\ix,0.04*\iy)[rotate=0,anchor=north]{{\scriptsize (b) Factor graph for control action update}};

    \node at (0.579*\ix,0.407*\iy)[rotate=0,anchor=north]{{\scriptsize $x_t$}};
    \node at (0.834*\ix,0.407*\iy)[rotate=0,anchor=north]{{\scriptsize $x_{t+1}$}};
    \node at (0.700*\ix,0.272*\iy)[rotate=0,anchor=north]{{\scriptsize $u_t$}};
    \node at (0.938*\ix,0.272*\iy)[rotate=0,anchor=north]{{\scriptsize $y_{t+1}$}};  
    \node at (0.825*\ix,0.160*\iy)[rotate=0,anchor=north]{{\scriptsize $k_{t+1}$}};
    \node at (0.695*\ix,0.49*\iy)[rotate=0,anchor=north]{{\scriptsize $f_{t+1}^x$}};
    \node at (0.86*\ix,0.335*\iy)[rotate=0,anchor=north]{{\scriptsize $f_{t+1}^y$}};
    \node at (0.65*\ix,0.17*\iy)[rotate=0,anchor=north]{{\scriptsize $f_{t}^{\text{goal},u}$}};
    \node at (0.95*\ix,0.12*\iy)[rotate=0,anchor=north]{{\scriptsize $f_{t+1}^{\text{goal},k}$}};
    \node at (0.888*\ix,0.56*\iy)[rotate=0,anchor=north]{{\scriptsize $f_{t+1}^{\text{goal},x}$}};

    \draw[gray!80, thin, ->] (0.81*\ix, 0.23*\iy) -- (0.81*\ix, 0.18*\iy);
    \draw[gray!80, thin, ->] (0.92*\ix, 0.10*\iy) -- (0.87*\ix, 0.10*\iy);

     \node at (0.755*\ix,0.04*\iy)[rotate=0,anchor=north]{{\scriptsize (c) Factor graph for sensing action update}};    

    \node at (0.28*\ix,0.34*\iy)[rotate=0,anchor=north]{{\scriptsize $\textcircled{1}$}};
    \node at (0.28*\ix,0.51*\iy)[rotate=0,anchor=north]{{\scriptsize $\textcircled{2}$}};
    \node at (0.25*\ix,0.48*\iy)[rotate=0,anchor=north]{{\scriptsize $\textcircled{3}$}};
    \node at (0.13*\ix,0.48*\iy)[rotate=0,anchor=north]{{\scriptsize $\textcircled{4}$}};
    \node at (0.23*\ix,0.39*\iy)[rotate=0,anchor=north]{{\scriptsize $\textcircled{5}$}};
    \node at (0.23*\ix,0.20*\iy)[rotate=0,anchor=north]{{\scriptsize $\textcircled{6}$}};

    \node at (0.35*\ix,0.995*\iy)[rotate=0,anchor=north]{{\scriptsize $\textcircled{1}$}};
    \node at (0.355*\ix,0.90*\iy)[rotate=0,anchor=north]{{\scriptsize $\textcircled{2}$}};
    \node at (0.45*\ix,0.995*\iy)[rotate=0,anchor=north]{{\scriptsize $\textcircled{3}$}};
    \node at (0.49*\ix,0.765*\iy)[rotate=0,anchor=north]{{\scriptsize $\textcircled{4}$}};
    \node at (0.575*\ix,0.775*\iy)[rotate=0,anchor=north]{{\scriptsize $\textcircled{5}$}};
    \node at (0.49*\ix,0.86*\iy)[rotate=0,anchor=north]{{\scriptsize $\textcircled{6}$}};

    
    \end{tikzpicture}
    \vspace{-3mm}
    \caption{Factor graph for the proposed AIF.}
    
    \label{fig:FactorGraph-action}
\end{figure}

To minimize the variational free energy of $q_1(x_{1:t})$, we adopt a factor-graph message passing scheme, as shown in Fig.~\ref{fig:FactorGraph-action}(a). The state transition and observation factors are defined as
\vspace{-2mm}
\begin{align}
f^x_t &= P(x_t \mid x_{t-1}, u_{t-1})
= \mathcal{N}(A x_{t-1} + B u_{t-1}, {Q}), \\
f^y_t &= P(y_t \mid x_t, k_t)
= \mathcal{N}(C x_t, {R}(k_t)),
\end{align}
with the prior belief $b(x_{t-1}) = \mathcal{N}(x_{t-1},m_{t-1}, W_{t-1})$, where $m_{t-1}$ and $W_{t-1}$ denote the mean and covariance from the previous step. By combining the transition messages $\textcircled{1}$–$\textcircled{3}$ with the observation messages $\textcircled{4}$–$\textcircled{6}$, the posterior belief over $x_t$ is obtained as $
b(x_t) \propto \mu_{f^x_t \rightarrow x_t}(x_t)
\cdot \mu_{f^y_t \rightarrow x_t}(x_t)$.
It results in a Gaussian belief $\mathcal{N}(x_t; m_t, W_t)$ with
\vspace{-3mm}
\begin{align}
W_t &= \big(W_{t|t-1}^{-1}
+ C^\top {R}^{-1}(k_t) C\big)^{-1}, \\
m_t &= W_t \big(
C^\top {R}^{-1}(k_t)y_t
+ W_{t|t-1}^{-1}m_{t|t-1}\big).
\end{align}
The message passing on the factor graph recovers the Kalman filtering recursion, showing that state estimation under active inference corresponds to minimizing the VFE.

\color{black}
 

\vspace{-3mm}
\subsection{Action Stage}
\vspace{-2mm}

In the action stage, the factor graph is augmented with goal-related factors that encode trajectory-tracking, control effort, and sensing cost. To avoid confusion with covariance matrices, we parameterize preferences by their covariance. The preference factors are  
\vspace{-3mm}
\begin{align}
    f^{\text{goal},x}_{t+1} &\propto 
    \exp\!\Big(-\tfrac12(x_{t+1}-x_{t+1}^{\text{desired}})^\top 
    Q_{\text{goal}} (x_{t+1}-x_{t+1}^{\text{desired}})\Big),  \nonumber \\
    f^{\text{goal},u}_t &\propto \exp\!\Big(-\tfrac12 u_t^\top R_{\text{goal}} u_t\Big), \nonumber \\
    f^{\text{goal},k}_{t+1} &\propto \exp\!\Big(-\tfrac12 \alpha k_{t+1}^2 -\tfrac12 \beta\log\det(R(k_{t+1}))\Big).  \nonumber
    \vspace{-3mm}
\end{align}
In the following, we derive the control action $u_t$ and the sensing action $k_{t+1}$ via message passing on the factor graph.  

\vspace{-3mm}
\subsubsection{Control Action Update}
\vspace{-2mm}

As illustrated in Fig.~\ref{fig:FactorGraph-action}(b), since future observations are unavailable, the update of $u_t$ relies on the current belief $b(x_t)$ from state estimation, the goal prior on $x_{t+1}$, and the control prior on $u_t$. 
The procedure is summarized as follows:
\vspace{-3mm}
\begin{align}
    \textcircled{1} & \mu_{f^y_{t+1} \to x_{t+1}}(x_{t+1}) \propto 1. \\
    \textcircled{2} & \mu_{f^{\text{goal},x}_{t+1} \to x_{t+1}}(x_{t+1}) 
    \propto \mathcal{N}(x_{t+1}; x_{t+1}^{\text{desired}}, Q_{\text{goal}}^{-1}). \\
    \textcircled{3} & \mu_{(x_{t+1}) \to f^x_{t+1}}
    (x_{t+1}) 
    \propto \mathcal{N}(x_{t+1}; x_{t+1}^{\text{desired}}, Q_{\text{goal}}^{-1}). \\
    \textcircled{4} & \mu_{b(x_t) \to f^x_{t+1}}(x_t) 
    \propto \mathcal{N}(x_t; m_t, W_t).\\
    \textcircled{5} & \mu_{f^x_{t+1} \to u_t}(u_t) \propto \mathcal{N}(u_t; m_{u_t}, W_{u_t}) \\
    \textcircled{6} & \mu_{f^{goal,u}_{t} \to u_t}(u_t) \propto \mathcal{N}(u_t; 0, R_{\text{goal}}^{-1}),  
\end{align}
where $\Sigma_{u_t} = AW_tA^\top + {Q} + Q_{\text{goal}}^{-1}$ with $W_{u_t} = (B^\top \Sigma_{u_t}^{-1} B)^{-1}$ and $m_{u_t} = W_{u_t} B^\top \Sigma_{u_t}^{-1} \big(x_{t+1}^{\text{desired}} - Am_t\big)$.
                   
The posterior belief over $u_t$ is obtained by multiplying the above message with the control preference factor: $b(u_t) \propto  \mu_{f^x_{t+1} \to u_t}(u_t) \cdot f^{\text{goal},u}_t(u_t)= \mathcal{N}(u_t; \tilde{m}_{u_t}, \tilde{W}_{u_t})$, 
where $\tilde{W}_{u_t}^{-1} = B^\top \Sigma_{u_t}^{-1} B + R_{\text{goal}}$ and $\tilde{m}_{u_t} = \tilde{W}_{u_t} B^\top \Sigma_{u_t}^{-1} \\ \big(x_{t+1}^{\text{desired}} - Am_t\big)$.
Finally, the control action is chosen as the mode of the posterior belief: $u_t^* = \arg\max_{u_t} b(u_t) = \tilde{m}_{u_t}$.

\vspace{-3mm}
\subsubsection{Sensing Action Update}
\vspace{-2mm}

After obtaining the control distribution $b(u_t)$, we update the sensing action $k_{t+1}$. 
The decision balances two effects: (i) allocating more subcarriers reduces the observation noise covariance $\bm R(k_{t+1})$ and improves state estimation; 
(ii) excessive allocation is penalized by the sensing cost prior.  

Since no future measurements are available at time $t+1$, the message from the likelihood factor $f^y_{t+1}$ to $k_{t+1}$ is uniform, i.e., $\mu_{f^y_{t+1}\to k_{t+1}}(k_{t+1}) \propto 1$.
Therefore, the update of $k_{t+1}$ is entirely driven by the sensing goal prior. 
With the sensing goal factor, the posterior distribution is $b(k_{t+1}) \propto f^{\text{goal},k}_{t+1}(k_{t+1}),$
and the optimal sensing action is obtained as
\vspace{-3mm}
\begin{equation}
    k_{t+1}^* = \arg\min_{k\in\{1,\dots,K\}}
    \Big\{\alpha{2}k^2 + \beta\log\!\det R(k)\Big\}.
    \vspace{-1mm}
\end{equation}

Although the optimal control $u_t^*$ and sensing action $k_{t+1}^*$ appear decoupled under the one-step strategy, they are inherently coupled in the closed-loop system. The control decision at time $t$ depends on the belief $b(x_t)$, which is shaped by past observations. In future work, this framework can be extended to multi-step strategies to capture such coupling more explicitly and to obtain more globally optimal solutions.

\color{black}




\vspace{-2mm}
\section{Numerical Results}
\vspace{-2mm}

The sensing and control interval of the UAV system is $\Delta t=0.1$~s, and the horizon $T=628$.
The process noise covariance is 
$  Q=\mathrm{diag}(10^{-2},10^{-3},10^{-2},10^{-3})$. 
The preference weights are 
$  Q_{\text{goal}}=\mathrm{diag}(10,1,0.1,0.1)$ and 
$  R_{\text{goal}}=\mathrm{diag}(0.01,0.01)$, with sensing cost coefficient $\alpha=10^{-3}$ and $\beta=1$. The sensing setup uses $K_{\text{tot}}=64$ subcarriers over $200$~MHz, and AOA error variance $\sigma_\theta^2=(5^\circ)^2$. The desired trajectory is given by $\rho_t=5\sin(0.5t)+50$~m, $v_{\rho,t}=5\cos(0.5t)$~m/s, 
and angular velocity $\omega=0.1$~rad/s. The initial state is set to $[100,\,0,\,5,\,0.1]^\top$.

\vspace{-2mm}
\subsection{Performance Comparison with Baselines}
\vspace{-2mm}

We consider two baseline schemes for comparison: 
\begin{itemize}
    \vspace{-3mm}

    \item \textbf{Greedy Control:} At each slot, the control input $u_t$ is obtained by directly steering the estimated state toward $x_{t+1}^{\text{desired}}$, while the sensing allocation $k_{t+1}$ follows the AIF result. 
    
     \vspace{-3mm}
    
    \item \textbf{Random Sensing:} At each slot, the control input $u_t$ is given by the AIF solution, whereas $k_{t+1}$ is randomly sampled from a Gaussian prior $k_{t+1} \sim \mathcal{N}(0, \alpha^{-1})$, which reflects the subcarrier allocation penalty.
    
     \vspace{-2mm}
\end{itemize}


\begin{table}[t]
\centering
\scriptsize 
\caption{Performance comparison of the proposed AIF and baselines.}
\begin{tabular}{lcccc}
\toprule
Method &  Control cost & Sensing cost & Total $J$ \\
\midrule
Proposed AIF & $3.74\times 10^3$ &   $-5.23\times 10^3$ & $-3.49\times 10^3$ \\
Greedy control   & $9.81\times 10^4$ &  $-5.22\times 10^3$ & $9.29\times 10^4$ \\
Random sensing  & $6.24\times 10^3$ & $5.11\times 10^2$ & $1.14\times 10^4$ \\
\bottomrule
\end{tabular}
\label{tab:performance}
\end{table}

\begin{figure}[t]
\vspace{-1mm}
\centering
\newcommand\ms{1.0}   
\newcommand\lw{1.2}   

\begin{tikzpicture}[font=\small, spy using outlines={rectangle, magnification=2.5, size=0.3cm, connect spies}]
\begin{axis}[%
    width=6cm,
    height=4cm,
    at={(0in,0in)},
    scale only axis,
    xmin=1e-4, xmax=1e-1,
    xmode=log,
    xlabel style={yshift=1ex},
    xlabel={$\alpha$},
    ymode=log,
    ymin=1e3, ymax=1e5,
    ylabel style={yshift=-0.5ex},
    ylabel={Control Cost},
    axis background/.style={fill=white},
    xmajorgrids, ymajorgrids,
    tick label style={font=\small},
    label style={font=\small},
    legend columns=1, 
    legend style={
        font=\scriptsize,
        text=black,
        at={(0.35,0.80)},
        anchor=south west,
        fill opacity=0.7,
        draw=white!15!black,
        legend cell align=left
    }
]
\addplot [color=blue, line width=\lw pt, mark=x, mark options={solid, blue}, mark size=\ms pt]
  table[row sep=crcr]{%
1e-4 3.36e3 \\
1e-3 3.74e3 \\
1e-2 5.98e3 \\
1e-1 2.71e4 \\
};
\end{axis}

\begin{axis}[%
    width=6cm,
    height=4cm,
    at={(0in,0in)},    
    scale only axis,
    xmin=1e-4, xmax=1e-1,
    xmode=log,
    ymin=0, ymax=70,
    ylabel style={yshift=0.5ex},
    ylabel={Average Subcarriers},
    axis y line*=right,
    axis x line=none,
    tick label style={font=\small},
    label style={font=\small},
    ymajorgrids=false,
    legend columns=1, 
    legend style={
        font=\scriptsize,
        text=black,
        at={(0.35,0.77)},
        anchor=south west,
        fill opacity=1,
        draw=black,
        legend cell align=left
    }
]

\addlegendimage{mark=x, color=blue, line width=\lw pt}
\addlegendentry{Control Cost}
\addplot [color=red, line width=\lw pt, mark=o, mark options={solid, red}, mark size=\ms pt]
  table[row sep=crcr]{%
1e-4 64 \\
1e-3 32 \\
1e-2 10 \\
1e-1 3 \\
};
\addlegendentry{Average number of subcarriers}

\end{axis}
\end{tikzpicture}
\vspace{-2mm}
\caption{Trade-off between control cost and sensing cost under different $\alpha$.}
\label{fig:Trade-off}
\end{figure}
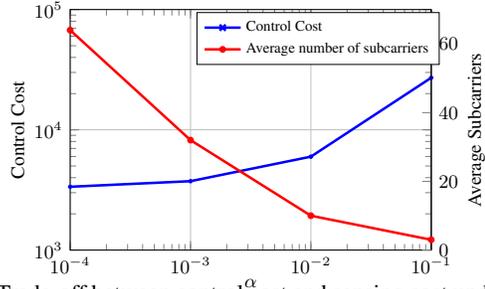

The numerical results are summarized in Table~\ref{tab:performance}. 
It can be observed that the proposed AIF framework, although relying only on one-step planning,  consistently outperforms both baselines. 
The greedy control method suffers from error accumulation due to process noise, as it greedily drives the state to $x_{t+1}^{\text{desired}}$ without uncertainty compensation. The random sensing method allocates sensing resources according to a random prior rather than task-oriented objectives, resulting in degraded performance. These results demonstrate that the proposed AIF-based joint optimization effectively enhances closed-loop SCC performance.

\vspace{-2mm}
\subsection{Trade-off between Control Cost and Sensing Cost}
\vspace{-2mm}

Fig.~\ref{fig:Trade-off} illustrates the trade-off between control and sensing costs as $\alpha$ varies. A larger $\alpha$ reduces the average number of subcarriers allocated to sensing, thereby lowering sensing cost but increasing estimation error. Consequently, more accurate state estimation with smaller sensing cost leads to lower control cost, highlighting the importance of jointly optimizing control and sensing. This result also confirms the effectiveness of the proposed AIF framework in balancing control performance and sensing resource consumption.

\vspace{-3mm}
\section{Conclusion}
\vspace{-3mm}

In this paper, we propose an AIF for closed-loop SCC. By constructing a unified generative model of UAV dynamics, sensing observations, and resource allocation, the SCC problem was formulated as variational and expected free energy minimization. Factor-graph message passing was employed for state estimation and action planning, enabling principled co-design of control and sensing. This work highlights the potential of AIF as a unified tool for intelligent wireless SCC system design.

\vspace{-2mm}
\section*{\normalsize  Acknowledgment}
\vspace{-3mm}
This work was supported in part by the National Academic Infrastructure for Supercomputing in Sweden (NAISS), by the SNS JU project 6G-DISAC under the EU's Horizon Europe research and innovation Program under Grant Agreement No 101139130, the Swedish Foundation for Strategic Research (SSF) (grant FUS21-0004, SAICOM), the ERANET CHIST-ERA Project MUSE-COM2, and in part by the Research Council of Finland (former Academy of Finland) Project Vision-Guided Wireless Communication.

\balance 
\bibliographystyle{IEEEtran}
\bibliography{IEEEabrv,ref}

\begin{thebibliography}{10}
\providecommand{\url}[1]{#1}
\csname url@samestyle\endcsname
\providecommand{\newblock}{\relax}
\providecommand{\bibinfo}[2]{#2}
\providecommand{\BIBentrySTDinterwordspacing}{\spaceskip=0pt\relax}
\providecommand{\BIBentryALTinterwordstretchfactor}{4}
\providecommand{\BIBentryALTinterwordspacing}{\spaceskip=\fontdimen2\font plus
\BIBentryALTinterwordstretchfactor\fontdimen3\font minus \fontdimen4\font\relax}
\providecommand{\BIBforeignlanguage}[2]{{%
\expandafter\ifx\csname l@#1\endcsname\relax
\typeout{** WARNING: IEEEtran.bst: No hyphenation pattern has been}%
\typeout{** loaded for the language `#1'. Using the pattern for}%
\typeout{** the default language instead.}%
\else
\language=\csname l@#1\endcsname
\fi
#2}}
\providecommand{\BIBdecl}{\relax}
\BIBdecl

\bibitem{Wang_AutonomousDriving_2019}
J.~Wang, J.~Liu, and N.~Kato, ``Networking and communications in autonomous driving: A survey,'' \emph{IEEE Commun. Surveys Tuts.}, vol.~21, no.~2, pp. 1243--1274, 2019.

\bibitem{Goal-Oriented_robot}
W.~Wu, Y.~Yang, Y.~Deng, and A.~Hamid~Aghvami, ``Goal-oriented semantic communications for robotic waypoint transmission: The value and age of information approach,'' \emph{IEEE Trans. Wireless Commun.}, vol.~23, no.~12, pp. 18\,903--18\,915, 2024.

\bibitem{Chang_IntegratedUAV_2022}
B.~Chang, W.~Tang, X.~Yan, X.~Tong, and Z.~Chen, ``Integrated scheduling of sensing, communication, and control for {mmWave/THz} communications in cellular connected {UAV} networks,'' \emph{IEEE J. Select. Areas Commun.}, vol.~40, no.~7, pp. 2103--2113, 2022.

\bibitem{li2024maximizing}
B.~Li, X.~Wang, and F.~Fang, ``Maximizing the value of service provisioning in multi-user {ISAC} systems through fairness guaranteed collaborative resource allocation,'' \emph{{IEEE} J. Sel. Areas Commun.}, vol.~42, no.~9, pp. 2243--2258, 2024.

\bibitem{li2025massive}
F.~Li and B.~Liao, ``Massive mimo-isac beamforming design via sensing energy maximization,'' in \emph{proc. IEEE ICASSP}.\hskip 1em plus 0.5em minus 0.4em\relax IEEE, 2025, pp. 1--5.

\bibitem{khalili2024efficient}
A.~Khalili, A.~Rezaei, D.~Xu, F.~Dressler, and R.~Schober, ``Efficient uav hovering, resource allocation, and trajectory design for {ISAC} with limited backhaul capacity,'' \emph{IEEE Trans. Wireless Commun.}, 2024.

\bibitem{zou2024energy}
J.~Zou, S.~Sun, C.~Masouros, Y.~Cui, Y.-F. Liu, and D.~W.~K. Ng, ``Energy-efficient beamforming design for integrated sensing and communications systems,'' \emph{IEEE Trans. Commun.}, vol.~72, no.~6, pp. 3766--3782, 2024.

\bibitem{duan2022distributed}
P.~Duan, L.~He, Z.~Duan, and L.~Shi, ``Distributed cooperative {LQR} design for multi-input linear systems,'' \emph{IEEE Trans. Control Netw. Syst.}, vol.~10, no.~2, pp. 680--692, 2022.

\bibitem{jin2025co}
H.~Jin, J.~Wu, W.~Yuan, F.~Liu, and Y.~Cui, ``Co-design of sensing, communications, and control for low-altitude wireless networks,'' \emph{IEEE Trans. Mobile Comput.}, no.~01, pp. 1--13, 2025.

\bibitem{meng2024communication}
Z.~Meng, D.~Ma, S.~Wang, Z.~Wei, and Z.~Feng, ``Modeling and design of the communication sensing and control coupled closed-loop industrial system,'' in \emph{Proc. IEEE GLOBECOM}, 2023, pp. 2626--2631.

\bibitem{fang2024sensing1}
X.~Fang, C.~Lei, W.~Feng, Y.~Chen, M.~Xiao, N.~Ge, and C.~Wang, ``Sensing-communication-computing-control closed-loop optimization for {6G} unmanned robotic systems,'' \emph{arXiv preprint arXiv:2410.18382}, 2024.

\bibitem{fang2025sensing2}
X.~Fang, C.~Lei, W.~Feng, Y.~Chen, M.~Xiao, N.~Ge, and C.-X. Wang, ``Sensing-communication-computing-control closed-loop optimization for {6G} digital twin-empowered unmanned robotic systems,'' \emph{{IEEE} J. Sel. Areas Commun.}, 2025.

\bibitem{pan2025rate}
G.~Pan, Z.~Li, A.~{\"O}z{\c{c}}elikkale, C.~H{\"a}ger, M.~F. Keskin, and H.~Wymeersch, ``Rate-limited closed-loop distributed {ISAC} systems: An autoencoder approach,'' \emph{arXiv preprint arXiv:2505.01780}, 2025.

\bibitem{10446575}
A.~Krayani, K.~Khan, L.~Marcenaro, M.~Marchese, and C.~Regazzoni, ``Self-supervised path planning in {UAV}-aided wireless networks based on active inference,'' in \emph{proc. IEEE ICASSP}, 2024, pp. 13\,181--13\,185.

\bibitem{obite2023active}
F.~Obite, A.~Krayani, A.~S. Alam, L.~Marcenaro, A.~Nallanathan, and C.~Regazzoni, ``Active inference for sum rate maximization in {UAV}-assisted cognitive noma networks,'' in \emph{Proc. IEEE WF-IoT}, 2023, pp. 1--6.

\bibitem{krayani2023goal}
A.~Krayani, K.~Khan, L.~Marcenaro, M.~Marchese, and C.~Regazzoni, ``A goal-directed trajectory planning using active inference in {UAV}-assisted wireless networks,'' \emph{Sensors}, vol.~23, no.~15, p. 6873, 2023.

\bibitem{otoshi2025coordinated}
T.~Otoshi and M.~Murata, ``Coordinated multi-point by distributed hierarchical active inference with sensor feedback,'' \emph{Computer Networks}, vol. 257, p. 110989, 2025.

\bibitem{gonzalez2024integrated}
N.~Gonz{\'a}lez-Prelcic, M.~F. Keskin, O.~Kaltiokallio, M.~Valkama, D.~Dardari, X.~Shen, Y.~Shen, M.~Bayraktar, and H.~Wymeersch, ``The integrated sensing and communication revolution for {6G}: Vision, techniques, and applications,'' \emph{Proceedings of the IEEE}, vol. 112, no.~7, pp. 676--723, 2024.

\bibitem{pan2025ai}
G.~Pan, Y.~Gao, Y.~Gao, Z.~Zhong, X.~Yang, X.~Guo, and S.~Xu, ``{AI}-driven wireless positioning: Fundamentals, standards, state-of-the-art, and challenges,'' \emph{arXiv preprint arXiv:2501.14970}, 2025.

\bibitem{TOA-CRLB}
S.-H. Kong and B.~Kim, ``Error analysis of the {OTDOA} from the resolved first arrival path in {LTE},'' \emph{IEEE Trans. Wireless Commun.}, vol.~15, no.~10, pp. 6598--6610, 2016.

\bibitem{AOA-CRLB}
C.-Y. Chen and W.-R. Wu, ``Joint {AoD}, {AoA}, and channel estimation for {MIMO-OFDM} systems,'' \emph{{IEEE} Trans. Veh. Technol.}, vol.~67, no.~7, pp. 5806--5820, 2018.

\bibitem{van2021application}
T.~van~de Laar, A.~{\"O}z{\c{c}}elikkale, and H.~Wymeersch, ``Application of the free energy principle to estimation and control,'' \emph{IEEE Trans. Signal Process.}, vol.~69, pp. 4234--4244, 2021.

\end{thebibliography}


\end{document}